\documentclass[12pt]{iopart}%
\usepackage{iopams}
\usepackage{cleveref}
\usepackage{graphicx}
\usepackage{setstack}
\usepackage{lmodern}
\usepackage{amssymb}%
\usepackage{amsfonts}%
\usepackage{graphicx}
\usepackage{cleveref}
\usepackage{amsfonts}
\usepackage{graphicx,dsfont}
\usepackage{color}
\usepackage{bbm}

\begin{document}
\title{New approach to periodic orbit theory of spectral
correlations}
\author{Petr Braun, Daniel Waltner}
\address{Fakult\"at f\"ur Physik, Universit\"at Duisburg-Essen, Lotharstra\ss e 1, 47048 Duisburg, Germany}
\begin{abstract}
 The existing periodic orbit theory of spectral correlations for classically chaotic
 systems relies on the Riemann-Siegel-like representation of the spectral determinants
 which is still largely hypothetical. We suggest a simpler derivation using analytic
 continuation of the periodic-orbit expansion of the pertinent generating function
 from the convergence border to physically important real values of its arguments.
 As examples we consider chaotic systems without time reversal as well as the Riemann
 zeta function and Dirichlet $L$-functions zeros.
\end{abstract}


\section{Introduction}

Application of the random matrix theory (RMT) to the statistics of the energy
spectra by Wigner and Dyson was originally motivated by the complexity of
spectra of heavy nuclei consisting of many strongly interacting particles.
However about 1980 it became clear that quantum systems with just two degrees
of freedom also have universal statistical spectral properties described by
RMT provided the corresponding classical motion is chaotic, for an overview
see Refs. \cite{Stockmann,Haake,Gutzwiller,Guhr}; this statement was
formulated as the famous Bohigas-Giannoni-Schmit conjecture \cite{BGS}. Its
understanding took  many investigations, to mention but a few
relevant here: the diagonal approximation \cite{Berry}, a heuristic derivation of the
oscillatory part of the correlation function for the unitary unversality class
\cite{BogK3}, the orbit partnership role in spectral correlations
\cite{Sieber}, the semiclassical treatment of spin in the level statistics \cite{Bolte, Heuslerspin, BolteHar}, the periodic orbit theory of the small time form factor \cite{MullerPhysRev}.  The complete semiclassical derivation of the two-point correlation
function for the unitary and orthogonal universality classes is given in
\cite{Heusler, Muller} and for the symplectic class in \cite{Braun}. Three- and higher-level
correlation functions are considered in \cite{NagaoMuller}.

The state-of-the-art calculation of spectral correlators is based on the
method of generating functions imported from the field theory of disorded
media, which are ratios of spectral determinants $\det\left(  E+e-H\right)$ 
with slightly different energy offsets $e$ averaged
over the central energy $E$. The fundamental role in the calculation is played
by the so called Riemann-Siegel lookalike \cite{Keatin,KeaMul}. This is the asymptotic representation
of a spectral determinant by an explicitly real sum over sets of periodic
orbits, or \textquotedblleft pseudo-orbits\textquotedblright, truncated at
pseudo-orbits with total period half the Heisenberg time $T_{H}$. The result is
far from obvious and has been proven only for some rare cases \cite{Wal,Har}, beyond the
Riemann-Siegel formula proper in the theory of the Riemann zeta function \cite{Tit,Edw,Ber500}. Its
usage in problems of spectral statistics looks also like an overkill since the
non-trivial boundary $T_{H}/2$ doesn't play any role; after the averaging,
summation over the pseudo-orbits is invariably extended to infinite periods.

\begin{figure}
 \raisebox{1cm}{\includegraphics[width=8cm]{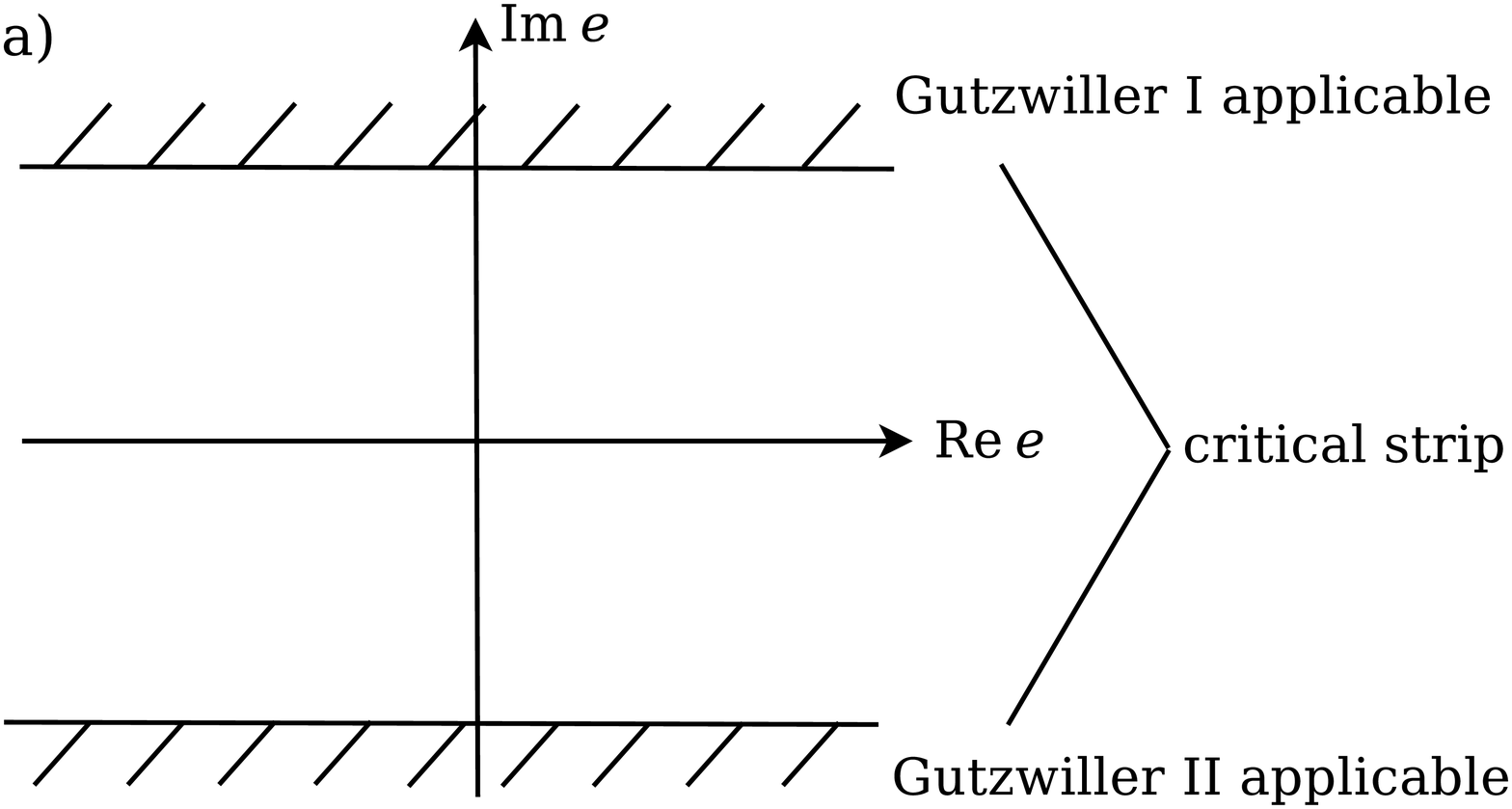}}\includegraphics[width=6cm]{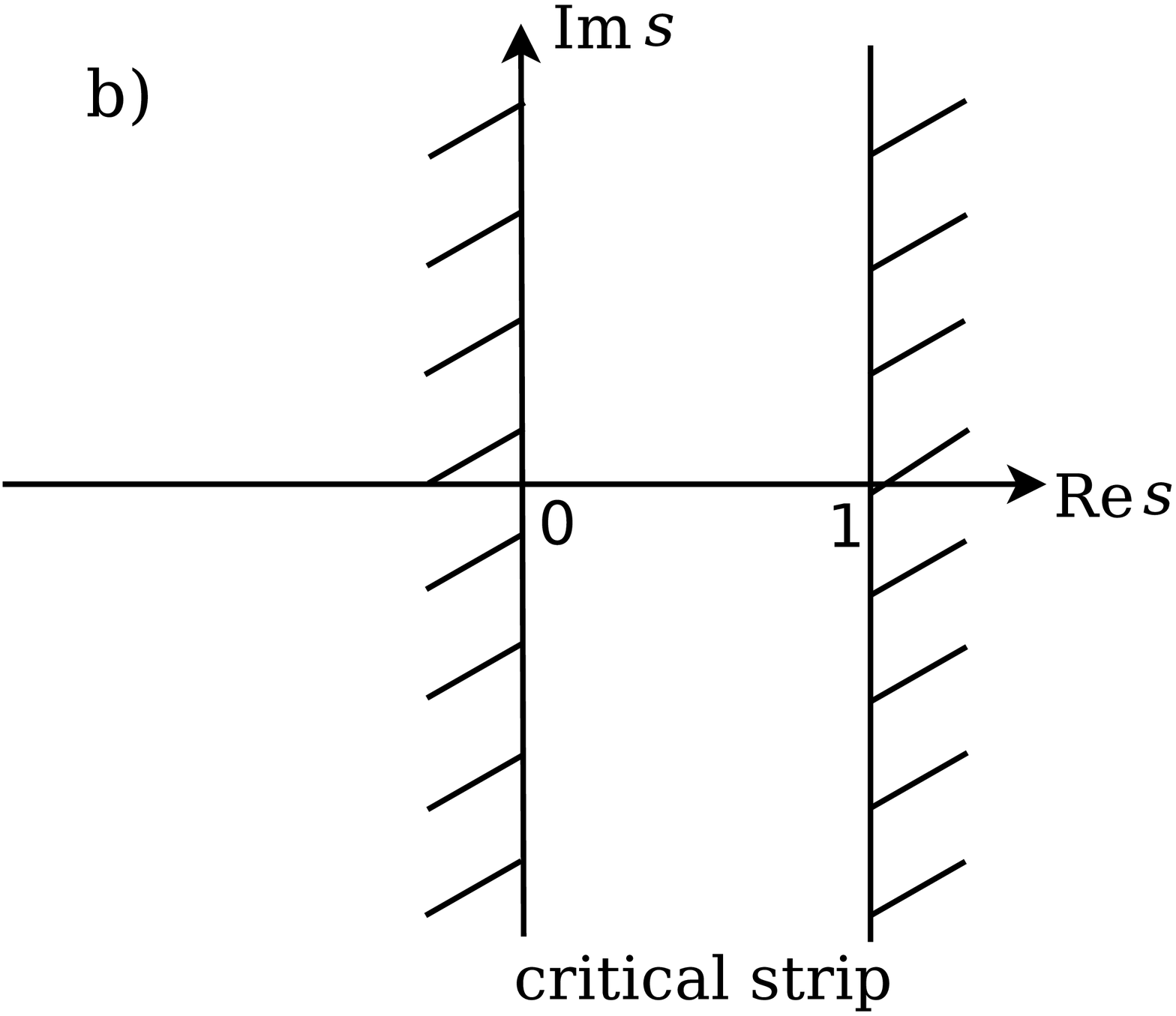}\label{fig}
\caption{Analogy between the semiclassical periodic orbit theory and the theory of the Riemann zeta function. a) The generating function:
two different Gutzwiller expansions are applicable outside the critical strip. b) The Riemann zeta function $\zeta(s)$ case.}
\end{figure}

Here we want to reformulate the theory in the form which avoids the use of the
Riemann-Siegel lookalike. The idea is to complexify the generating function by
letting its energy arguments have an imaginary part large enough for
convergence of the pertinent Gutzwiller expansions like the Gutzwiller trace formula. It can be done in two ways
depending on the sign of the imaginary parts of the energies $e$ which, after averaging, provides
two different semiclassical asymptotics which can be analytically continued inside the critical strip. We will refer to these two ways as I and II, for an illustration, see Fig.\ \ref{fig} a).
The semiclassical approximation of
the generating function in the vicinity of the real energy axis smoothly
matching both of these asymptotics is their sum since the \textquotedblleft
wrong\textquotedblright\ component is exponentially subdominant for energies
with sufficiently large absolute value of the imaginary part.

The paper is built in the following way. In Section 2 we describe the
semiclassical periodic orbit theory of the generating function justifying
its composition of two parts responsible for the monotonic and oscillatory
parts of the level-level correlation function. In Section 3 we remind how the
diagonal approximation neglecting the orbit action correlation is introduced
for the generating functions; unlike Berry's diagonal approximation it
provides both the oscillatory and monotonic parts of the spectral correlator. In
Section 4 we show that our approach applied to the spectral correlator of
the Riemann zeta function $\zeta(s)$ and Dirichlet $L-$functions zeros reproduces in a more
transparent way the results of Bogomolny and Keating \cite{BogK3} and Bogomolny \cite{Bog}. 
We apply here arguments similar to the ones used in the case of the periodic orbit theory, for an illustration see Fig.\ \ref{fig}b). In
Section 5 we return to dynamical systems without time-reversal symmetry and show how the oscillatory part of the
correlator including nonuniversal effects stemming from repetitions can be obtained by our approach.
In section 6 we explain why in the presence of time-reversal symmetry
the oscillatory part of the correlator cannot
be obtained without taking into account periodic orbit action
correlations explicitly. Finally we conclude in section 7.

\section{Correlation function}

The level-level correlation function can be defined as%
\[
R\left(  e\right)  =\frac{\left\langle \rho\left(  E-\frac{e}{2}\right)
\rho\left(  E+\frac{e}{2}\right)  \right\rangle _{E}}{\left\langle \rho\left(
E-\frac{e}{2}\right)  \right\rangle _{E}\left\langle \rho\left(  E+\frac{e}%
{2}\right)  \right\rangle _{E}}-1
\]
where $\rho\left(  E\right)  $ is the level density; averaging is done over an
interval of the reference energy $E$ large compared with the mean level
spacing $\delta=1/\bar{\rho},\quad\bar{\rho}=\left\langle \rho\right\rangle
_{E},$ but small in classical terms; $e$ is the energy offset. The non-trivial part of $R(e)$
\begin{equation}
r\left(  e\right)  \equiv\left\langle \rho\left(  E-\frac{e}{2}\right)
\rho\left(  E+\frac{e}{2}\right)  \right\rangle _{E}\label{paircorrelation}%
\end{equation}
to which we shall refer below as the correlation function for brevity, can be
expressed in terms of the averaged ratio of four spectral determinants, or the
generating function,%
\[
Z\left(  e_{A},e_{B},e_{C},e_{D}\right)  =\left\langle \frac{\det\left(
E+e_{C}-H\right)  \det\left(  E-e_{D}-H\right)  }{\det\left(  E+e_{A}%
-H\right)  \det\left(  E-e_{B}-H\right)  }\right\rangle _{E}.
\]
The connection between these quantities is given by the formula,%
\begin{equation}
r(e)=-\frac{1}{2\pi^{2}}\mathrm{Re}\left.  \frac{\partial^{2}Z}{\partial
e_{A}\partial e_{B}}\right\vert _{\parallel},\label{ZtoR}%
\end{equation}
assuming that the arguments of $Z$ have imaginary parts of the same sign for
$e_{A},e_{B}$ as well as for $e_{C},e_{D}$: otherwise the result of averaging with respect to $E$ would be zero. After the averaging, all
$e_{X},\quad X=A,B,C,D,$ must be set to $e/2$ which is indicated by $\left(
\parallel\right)  $ in (\ref{ZtoR}); this is the so called \textquotedblleft
columnwise\textquotedblright\ limit of \cite{Heusler}.

Without loss of generality we can fix for example ${\rm{Im}}\,e_{A,B}>0$ with the limit value $0_{+}$. As
regards the energy offsets in the numerator we can then choose either
${\rm{Im}}\,e_{C,D}>0$ or ${\rm{Im}}\,e_{C,D}<0$ with the limit
values $0_{+}$ and $0_{-}$ respectively. The two corresponding limits of $Z$
must coincide as the latter quantity is well defined for real $e_C$ and $e_D$. 
On the other hand, changing the sign of ${\rm{Im}}\,%
e_{C,D}$ is equivalent to the replacement $e_{C}\rightleftarrows-e_{D}$ in $Z$. We
come thus to the so called Weyl symmetry relation valid for real $e_{C,D}$,%
\begin{equation*}
Z\left(  e_{A},e_{B},e_{C},e_{D}\right)    =Z\left(  e_{A},e_{B}%
,-e_{D},-e_{C}\right)\hspace*{1cm} {\rm{with}}\hspace*{1cm}
{\rm{Im}}\,e_{C,D}  =0\nonumber .
\end{equation*}

The semiclassical asymptotics of the generating function follows from the
Gutz\-willer periodic orbit expansion \cite{Gutzwiller} of the trace of the resolvent. It
converges when ${\rm{Im}}\,E$ is positive and large enough (the value
guaranteeing absolute convergence is half the Kolmogorov-Sinai entropy, in
view of the exponential orbit proliferation) and is given by
\begin{equation}
{\rm{Tr}}\left(  E-H\right)  ^{-1}   \sim-i\pi\bar{\rho}\left(
E\right)  -\frac{i}{\hbar}\sum_{p}T_{p}F_{p}\mathrm{e}^{iS_{p}\left(
E\right)  /\hbar}\label{TrResolvIMgt0}\hspace*{6mm}
{\rm{for}}\hspace*{6mm}{\rm{Im}}\,E  >0,
\end{equation}
where we neglected the contributions from the orbit repetitions. The classical
action of the orbit $p$ is denoted by $S_p$, the duration of the periodic orbits $T_p$
and the stability prefactor by $F_{p}$, for their precise form see Ref.\ \cite{Gutzwiller}.
The expansion of the spectral determinant then follows from the identity
$\det\left(  E-H\right)  =\exp{\rm{Tr}}\ln(E-H)$ which gives,%
\begin{eqnarray}
&\det\left(  E-H\right)\propto\exp{\rm{Tr}}\int^{E}dE^{\prime
}\left(  E^{\prime}-H\right)  ^{-1}\label{Positive}\propto\exp\left[  -i\pi\bar{N}\left(  E\right)
\right]\prod_pd_p(E)  ,\nonumber\\ &{\rm{with}}\hspace*{6mm}
{\rm{Im}}E   >0\hspace*{6mm} {\rm{and}}\hspace*{6mm} d_p(E)  =\exp%
\left[-F_{p}\mathrm{e}^{iS_{p}\left(  E\right)  /\hbar}\right],
\end{eqnarray}
where%
\begin{equation}\label{count}
\bar{N}\left(  E\right)  =\int^{E}dE^{\prime}\bar{\rho}\left(  E^{\prime
}\right)
\end{equation}
denotes the averaged level counting function.
The spectral determinant with
the argument whose imaginary part is negative, is obtained by complex
conjugation,
\begin{equation}
\det\left(  E-H\right)    \propto\exp\left[  i\pi\bar{N}\left(  E\right)
  \right] \prod_pd_p^\ast(E) \label{neggamma}\hspace*{6mm}
{\rm{with}}\hspace*{6mm}{\rm{Im}}E <0.\nonumber
\end{equation}
The two expansions can be used to get the semiclassical approximation of the
generating function. Considering that ${\rm{Im}}\,e_{A,B}>0$ we
substitute (\ref{Positive}) for $\det(E+e_{A}-H)$ and (\ref{neggamma}) for
$\det\left(  E-e_{B}-H\right)  $. As regards the numerator there are two
possibilities. If we choose ${\rm{Im}}\,e_{C,D}>0$ we must use
(\ref{Positive}) for $\det(E+e_{C}-H)$ and (\ref{neggamma}) for $\det
(E-e_{D}-H)$. Expanding $\bar{N}\left(  E+e_{A}\right)  =\bar{N}\left(
E\right)  +e_{A}\bar{\rho}$ etc. we obtain the semiclassical asymptotics of
the determinants ratio as
\begin{eqnarray}
Z  & \sim Z^{I}\left(  e_{A,}e_{B},e\,_{C},e_{D}\right)  =\mathrm{e}%
^{i\pi\left(  e_{A}+e_{B}-e_{C}-e_{D}\right)  \bar{\rho}}\left\langle\prod_pz_p(e_{A,}e_{B},e_{C},e_{D})\right\rangle
,\nonumber\\ &\hspace*{6mm}{\rm{with}}\hspace*{6mm}
{\rm{Im}}\,e_{C,D}   >0 \hspace*{6mm}{\rm{and}}\hspace*{6mm} {\rm{Im}}\,e_{A,B}>0\nonumber
\end{eqnarray}
with the definition
\begin{equation}
z_p(e_{A,}e_{B},e\,_{C},e_{D})=d_p(E+e_{C})d_p^\ast(E-e_{D})d_p(E+e_{A})d_p^\ast(E-e_{B}).
\end{equation}
If we choose ${\rm{Im}}\,e_{C,D}<0$ we get an alternative expression,
\begin{eqnarray}
Z  & \sim Z^{II}\left(  e_{A,}e_{B},e\,_{C},e_{D}\right)  =Z^{I}\left(
e_{A,}e_{B},-e\,_{D},-e_{C}\right)  =\mathrm{e}^{i\pi\left(  e_{A}+e_{B}%
+e_{C}+e_{D}\right)  \bar{\rho}}\nonumber\\
& \times\left\langle\prod_pz_p(e_{A,}e_{B},-e_{D},-e_{C}) \right\rangle
\hspace*{4mm}{\rm{with}}\hspace*{4mm}
{\rm{Im}}\,e_{C,D} <0\hspace*{4mm}{\rm{and}}\hspace*{4mm}{\rm{Im}}\,e_{A,B}>0\nonumber.
\end{eqnarray}
The functions $Z^{I},Z^{II}$ resulting after the energy averaging are
well-behaved analytic functions of their arguments in which ${\rm{Im}}\,%
e_{A,B}$ can be brought to the limit $0_{+}$ whereas $e_{C,D}$ can take any
complex value. This is motivated by the fact that the presence of imaginary
parts of the energy arguments in the denominator is obligatory in order to avoid the poles;
on the other hand $e_C$ and $e_D$ can take any real values.
It is important that due to the prefactors $\exp[\mp
i\pi\left(  e_{C}+e_{D}\right)  \bar{\rho}]$ the function $Z^{II}\left(
e_{A,}e_{B},e\,_{C},e_{D}\right)  $ analytically continued to the
\textquotedblleft alien\textquotedblright\ domain with ${\rm{Im}}\,%
e_{C,D}$ large and positive, is exponentially smaller than $Z^{I}\left(
e_{A,}e_{B},e\,_{C},e_{D}\right)  $. Similarly the analytic continuation of
$Z^{I}$ to ${\rm{Im}}\,e_{C},e_{D}<0$ is exponentially small compared to
$Z^{II} $. With $e_{C,D}$ in the vicinity of the real axis $Z^{I},Z^{II}$ are
of the same order of magnitude.

What we are interested in is the semiclassical asymptotics of the generating
function for $e_{C,D}$ real. We know that its analytical continuation to
$e_{C,D}$ complex tends to $Z^{I}$ for large ${\rm{Im}}\,e_{C,D}>0$, and
to $Z^{II}$ for ${\rm{Im}}\,e_{C,D}<0$. Their sum has this property and yields
thus the semiclassical asymptotic for $Z$,
\[
Z\sim Z_{\rm{sc}}=Z^{I}+Z^{II}
\]
which uniformly approximates $Z$ for both signs of the imaginary parts and
reduces either to $Z^{I}$ or to $Z^{II}$ when $\left\vert {\rm{Im}}\,%
e_{C,D}\right\vert $ is large. Like the exact generating function it obeys the
Weyl symmetry relation when $e_{C,D}$ are real,
\[
Z_{\rm{sc}}\left(  e_{A},e_{B},e_{C},e_{D}\right)  =Z_{\rm{sc}}\left(  e_{A}%
,e_{B},-e_{D},-e_{C}\right)  .
\]

\section{Diagonal approximation}\label{diagappro}

Consider $Z^{I}$ with all its arguments containing large positive imaginary
parts safeguarding convergence of the Gutzwiller expansions. Then $Z^{I}$ can
be written as the energy average of a converging product over periodic orbits.
Expanding the classical actions like $S_{p}\left(  E+e_{A}\right)
=S_{p}\left(  E\right)  +T_{p}e_{A}$ etc. we can write it as,%
\[
Z^{I}\left(  e_{A,}e_{B},e\,_{C},e_{D}\right)  =\mathrm{e}^{i\pi\left(
e_{A}+e_{B}-e_{C}-e_{D}\right)  \bar{\rho}}\left\langle \prod_{p}%
z_{p}\right\rangle _{E},
\]
where%
\[
\fl z_{p}=\exp\left[  F_{p}\mathrm{e}^{iS_{p}\left(  E\right)  /\hbar}\left(
\mathrm{e}^{iT_{p}e_{A}/\hbar}-\mathrm{e}^{iT_{p}e_{C}/\hbar}\right)
+F_{p}^{\ast}\mathrm{e}^{-iS_{p}\left(  E\right)  /\hbar}\left(
e^{iT_{p}e_{B}/\hbar}-e^{iT_{p}e_{D}/\hbar}\right)  \right].
\]
Neglecting correlations between $z_{p}$ corresponding to orbits which do not
have exactly the same action we can replace the average of the product by the
product of averages. In the absence of time reversal and spatial symmetry the
orbit actions are non-degenerate. Disregarding the periodic orbit repetitions
we get,%
\begin{equation*}
Z^{I}  \approx\mathrm{e}^{i\pi\left(  e_{A}+e_{B}-e_{C}-e_{D}\right)
\bar{\rho}}\prod_{p}\left\langle z_{p}\right\rangle.
\end{equation*}

By the analogous steps as in \cite{Heusler} we obtain,%
\begin{equation*}
Z^{I}   \approx\mathrm{e}^{i\pi\left(  e_{A}+e_{B}-e_{C}-e_{D}\right)
\bar{\rho}}\,\tilde{Z}_{GUE}^{I}\hspace*{6mm}{\rm{with}}\hspace*{6mm}
\tilde{Z}_{GUE}^{I}  =\frac{\left(  e_{A}+e_{D}\right)  \left(  e_{B}%
+e_{C}\right)  }{\left(  e_{A}+e_{B}\right)  \left(  e_{C}+e_{D}\right)  }.
\end{equation*}
The second component of $Z$ is obtained by the interchange of $e_{C,D}$ and
change of their signs (the Weyl symmetry operation),%
\begin{equation*}
Z^{II}   \approx\mathrm{e}^{i\pi\left(  e_{A}+e_{B}+e_{C}+e_{D}\right)
\bar{\rho}}\tilde{Z}_{GUE}^{II}\hspace*{6mm}{\rm{with}}\hspace*{6mm}
\tilde{Z}_{GUE}^{II}   =-\frac{\left(  e_{A}-e_{C}\right)  \left(
e_{B}-e_{D}\right)  }{\left(  e_{A}+e_{B}\right)  \left(  e_{C}+e_{D}\right)
}.
\end{equation*}
The correlation function obtained as the real part of $\left.  \partial
_{e_{A}e_{B}}^{2}\left(  Z^{I}+Z^{II}\right)  \right\vert _{\parallel}$
coincides with the correlation function of the GUE ensemble, its monotonic and
oscillatory parts produced by $Z^{I}$ and $Z^{II},$ respectively. We observe that
the product $\left(  e_{A}-e_{C}\right)  \left(  e_{B}-e_{D}\right)  $ in
$\tilde{Z}_{GUE}^{II}$ turns into zero in the columnwise limit $e_{A,B,C,D}\to e/2$ such that both
differentiations must be spent on it,\thinspace\ hence $\left(  \partial
_{e_{A}e_{B}}^{2}Z^{II}\right)  _{\parallel}=-\mathrm{e}^{i2\pi e\bar{\rho}%
}/e^{2}$.

The usual diagonal approximation as introduced by Berry \cite{Berry} neglects terms corresponding to
different orbits in the expansion of the correlation function itself. Only
the non-oscillatory part of the correlator is then recovered; the oscillatory
part is regarded as the result of the periodic orbits action correlations. It
was demonstrated by Keating \cite{Keating} and Bogomolny \cite{Bog} that the oscillatory contribution to
the correlation function of the Riemann zeta zeros results from the tendency of the prime
numbers (``periodic orbits" of the problem) to avoid one another
described by the Hardy-Littlewood conjecture. The respective orbit correlation
mechanism \cite{Argaman} in physical systems without time reversal may thus be called ``the
Hardy-Littlewood lookalike"; its nature is still not quite clear. It is
remarkable that we can completely neglect that correlation using instead the
diagonal approximation for the Weyl-symmetric generating function, and get the
same result. That equivalence seems to hold also for corrections to the universal
RMT result; compare the Riemann zeta treatment by Bogomolny \cite{Bog}, Conrey and Snaith \cite{Conrey}
and the one below in this paper.

\section{Zeros of Riemann zeta function}

Following the idea of Hilbert and P\'{o}lya, the non-trivial zeros of the
Riemann zeta function may be interpreted as the eigenvalues of a self-adjoint
Hamilton operator of a virtual dynamical system. The analogy between the
density of zeros of the Riemann zeta and the Gutzwiller trace formula for
chaotic quantum systems provides further support for this spectral
interpretation. Another evidence for this interpretation provides the behavior
of the pair correlation of the zeros of the Riemann zeta function first
studied by Montgomery \cite{Montgomery}; they are the same as for the
eigenvalues of the Gaussian unitary ensemble of random matrices (GUE) as
confirmed by Odlyzko's numerical calculations \cite{Odlyzko}.

The spectral properties of the Riemann zeta function and closely related
$L$-functions were a subject of an innumerable number of investigations; in
connection with the present paper we would like to mention the works of
Keating \cite{Keating}, Bogomolny et al.
\cite{BogK3,Bog,BogK1,BogK2,BogK4,BogK5} and Berry et al. \cite{BerK}.
 See also Ref.\ \cite{Conrey} for a review and the references
therein. Here we show that the pair correlation function including the
system-specific corrections to the GUE prediction can easily be obtained by
the method described above.

\subsection{The generating function}

The Riemann zeta function can be defined by the Euler product over primes,%
\[
\zeta\left(  s\right)  =\prod_{p}\frac{1}{1-p^{-s}}
\]
which converges provided ${\rm{Re}}\,s>1$ but can be analytically
continued to the whole complex plane. According to the Riemann hypothesis, all
non-trivial zeros of zeta have the form $s_{k}=\frac{1}{2}+iE_{k}$ with
$E_{k}$ real; the spectrum is symmetric with respect to $E=0$. The
corresponding spectral determinant $\Delta(E)=\prod_{k}A\left(  E,E_{k}%
\right)  (E-E_{k})$ where $A\left(  E,E_{k}\right)  $ stands for a suitable
regularizing function, is related to the Riemann zeta function by
\begin{equation}
\Delta(E)=B(E)e^{-i\pi\bar{N}(E)}\zeta\left(  \frac{1}{2}-iE\right)
\label{SpecDetDef}%
\end{equation}
with $\bar{N}(E)$ being the mean counting function, i.e.\ the integral of the
mean zeros density $\bar{\rho}\left(  E\right)  $ with respect to $E$, and
$B(E)$ a smooth function positive for real $E$ whose explicit form depends on
the regularization choice \cite{Keating}. We need only the asymptotic expressions for large
$E$:%
\[
\bar{N}\left(  E\right)  \sim\frac{E}{2\pi}\left(\ln\frac{E}{2\pi}-1\right),\quad\bar{\rho
}\left(  E\right)  =\frac{d\bar{N}}{dE}\sim\frac{1}{2\pi}\ln\frac{E}{2\pi}.
\]
The Euler product for the spectral determinant converges for
${\rm{Im}}\,E>1/2, $ and then%
\begin{equation}
\Delta\left(  E\right)  =B\left(  E\right)  \mathrm{e}^{-i\pi\bar{N}\left(
E\right)  }\prod_{p}\frac{1}{1-p^{-1/2+iE}}.\label{impos}%
\end{equation}
When ${\rm{Im}}\,E<-\frac{1}{2}$ the complex conjugation gives
\begin{equation}
\Delta\left(  E\right)  =B\left(  E\right)  \mathrm{e}^{i\pi\bar
{N}\left(  E\right)  }\prod_{p}\frac{1}{1-p^{-1/2-iE}}.\label{imneg}%
\end{equation}
Calculating $\frac{d}{dE}\log\Delta\left(  E\right)  $ we would get the
Gutzwiller-like formula for the trace of the virtual resolvent with primes as
periodic orbits, analogous to (\ref{TrResolvIMgt0}) and its complex conjugate
but taking into account the orbit repetitions. In fact we don't need that to 
construct the generating function since the expansion for 
\begin{equation*}
Z   =\left\langle \frac{\Delta\left(  E+e_{C}\right)  \Delta\left(
E-e_{D}\right)  }{\Delta\left(  E+e_{A}\right)  \Delta\left(  E-e_{B}\right)
}\right\rangle _{E}\hspace*{6mm}{\rm{with}}\,
E\,{\rm{ real}},\,{\rm{Im}}\,e_{A,B} >0,
\end{equation*}
follows directly from the Euler product of $\zeta$. The generating function can be written
in two versions. If ${\rm{Im}}\,e_{C,D}>0$ we use (\ref{impos}) for
$\Delta\left(  E+e_{C}\right)  ,\Delta\left(  E+e_{A}\right)  $ and
(\ref{imneg}) for the two other determinants. Let us assume all deviations $e$
small compared with $E\gg1$ such that $\bar{N}\left(  E+e_{A}\right)
\approx\bar{N}\left(  E\right)  +\bar{\rho}\,e_{A}$ etc. and
$\left|{\rm{Im}}\,e_{X}\right|>1/2,\quad X=A,B,C,D$. Denoting%
\[
a_{p}\equiv\frac{\mathrm{e}^{ie_{A}\ln p}}{\sqrt{p}}
\]
and similar for $b_{p},c_{p},d_{p}$ we have for ${\rm{Im}}\,e_{A,B,C,D}>1/2$,%
\begin{eqnarray}
Z^{I}\left(  e_{A},e_{B},e_{C},e_{D}\right)   & =\exp\left[  -i\pi\left(
e_{C}+e_{D}-e_{A}-e_{B}\right)  \bar{\rho}\right]  \left\langle \prod_{p}%
z_{p}\right\rangle _{E},\nonumber\\ \nonumber \hspace*{6mm}{\rm{with}}\hspace*{16mm}
z_{p} & =\frac{\left(  1-a_{p}\mathrm{e}^{iE\ln p}\right)  \left(
1-b_{p}\mathrm{e}^{-iE\ln p}\right)  }{\left(  1-c_{p}\mathrm{e}^{iE\ln
p}\right)  \left(  1-d_{p}\mathrm{e}^{-iE\ln p}\right)  },\nonumber
\end{eqnarray}
we neglected the $e$-dependence of the smooth function $B$. The alternative
representation $Z^{II}$ with ${\rm{Im}}\,e_{C,D}$ negative is obtained
from $Z^{I}$ by the Weyl substitution,%
\begin{equation*}
\fl Z^{II}\left(  e_{A},e_{B},e_{C},e_{D}\right)    =Z^{I}\left(  e_{A}%
,e_{B},-e_{D},-e_{C}\right) \hspace*{4mm}{\rm{with}}\hspace*{4mm}
{\rm{Im}}\,e_{A,B}  >1/2,\quad{\rm{Im}}\,e_{C,D}<-1/2;
\end{equation*}
with the substitution rule $\left\{  c_{p},d_{p}\right\}  \rightarrow
\left\{  \left(  d_{p}p\right)  ^{-1},\left(  c_{p}p\right)  ^{-1}\right\}  $.

Now we introduce the diagonal approximation by the assumption that $z_{p}$
associated with different primes are uncorrelated such that the energy average
of the product of $z_{p}$ can be replaced by the product of their averages
writing
\[
Z^{I}\approx\exp\left[  -i\pi\left(  e_{C}+e_{D}-e_{A}-e_{B}\right)  \bar
{\rho}\right]  \prod_{p}\left\langle z_{p}\right\rangle _{E}.
\]
The factors in the product are elementarily calculated,%
\begin{eqnarray}
\left\langle z_{p}\right\rangle _{E}  & =\frac{\ln p}{2\pi}\int_{0}%
^{\frac{2\pi}{\ln p}}dE\,z_{p}=\frac{1}{2\pi i}%
{\displaystyle\oint_{\left\vert u\right\vert =1}}
\frac{du}{u}\frac{\left(  1-a_{p}u\right)  \left(  1-b_{p}/u\right)  }{\left(
1-c_{p}u\right)  \left(  1-d_{p}/u\right)  }\label{zpav}\nonumber\\
& =1+\frac{\left(  b_{p}-d_{p}\right)  \left(  a_{p}-c_{p}\right)  }%
{1-c_{p}d_{p}}.
\end{eqnarray}
Considering that $\zeta\left(  1-i\left(  e_{A}+e_{B}\right)  \right)
=\prod_{p}\left(  1-a_{p}b_{p}\right)  ^{-1}$ etc, we write $Z^{I}=Z_{\zeta
}^{I}\,\Phi^{I}$, where
\begin{equation}
\fl Z_{\zeta}^{I}=\exp\left[  -i\pi\left(  e_{C}+e_{D}-e_{A}-e_{B}\right)
\bar{\rho}\right]  \frac{\zeta\left(  1-i\left(  e_{A}+e_{B}\right)  \right)
\zeta\left(  1-i\left(  e_{C}+e_{D}\right)  \right)  }{\zeta\left(  1-i\left(
e_{C}+e_{B}\right)  \right)  \zeta\left(  1-i\left(  e_{A}+e_{D}\right)
\right)  }.\label{zetz}%
\end{equation}
The function $\Phi^{I}$ is a product over primes,%
\begin{eqnarray}
\fl\Phi^{I}\left(  e_{A},e_{B},e_{C},e_{D}\right)   & =\prod_{p}\phi_{p}%
^{I},\quad\label{PhiI}\\ \fl \hspace*{16mm}{\rm{with}}\hspace*{6mm}
\phi_{p}^{I}  & =\frac{(1-a_{p}b_{p})(1-c_{p}d_{p})}{(1-b_{p}c_{p}%
)(1-a_{p}d_{p})}\left\langle z_{p}\right\rangle =1-\frac{a_{p}b_{p}%
(a_{p}-c_{p})(b_{p}-d_{p})}{(1-b_{p}c_{p})(1-a_{p}d_{p})}.\label{phip}%
\end{eqnarray}
The second component $Z^{II}$ is obtained by the Weyl substitution
$e_{C}\rightleftarrows-e_{D}$ and can be written as $Z_{\zeta}^{II}\Phi^{II}$
with
\[
\fl Z_{\zeta}^{II}=\exp\left[  -i\pi\left(  -e_{C}-e_{D}-e_{A}-e_{B}\right)
\bar{\rho}\right]  \frac{\zeta\left(  1-i\left(  e_{A}+e_{B}\right)  \right)
\zeta\left(  1+i\left(  e_{C}+e_{D}\right)  \right)  }{\zeta\left(  1-i\left(
e_{B}-e_{D}\right)  \right)  \zeta\left(  1-i\left(  e_{A}-e_{C}\right)
\right)  }
\]
and $\Phi^{II}\left(  e_{A},e_{B},e_{C},e_{D}\right)  =\Phi^{I}\left(
e_{A},e_{B},-e_{D},-e_{C}\right)  .$

Considering the exponential
factors in $Z_{\zeta}$ the sum%
\begin{equation}
Z\sim Z^{I}+Z^{II}\label{z1plusz2}%
\end{equation}
will smoothly interpolate between $Z^{I}$ and $Z^{II}$ as we change
${\rm{Im}}\,e_{C,D}$ from positive to negative values, providing thus
the high-energy asymptotics of the generating function for real $e_{C,D}$.
It coincides with the result obtained differently in \cite{Conrey}.

\subsection{The correlation function}

The correlation function based on Eq.\ (\ref{z1plusz2}) consists of two summands.
Differentiation of $Z^{I}$ with the subsequent columnwise identification $(||)
$ $e_{A}=e_{B}=e_{C}=e_{D}=e/2$ gives the contribution which Bogomolny
\cite{Bog} refers as the diagonal one. It is easily checked that
\begin{equation}
\left(  Z_{\zeta}^{I}\right)  _{||}=1,\hspace*{1cm}\left(  \frac{\partial
Z^{I}}{\partial e_{A}}\right)  _{||}=\left(  \frac{\partial Z^{I}}{\partial
e_{B}}\right)  _{||}=0\label{pa1}%
\end{equation}
and for the second derivative
\begin{equation}
\left(  \frac{\partial^{2}Z_{\zeta}^{I}}{\partial e_{A}\partial e_{B}}\right)
_{||}=\frac{\partial^{2}\ln\zeta(1-ie)}{\partial e^{2}}=\frac{1}{e^{2}%
}+(\gamma_{0}^{2}+2\gamma_{1})+O(e).\label{pa2}%
\end{equation}
For $\Phi^{I}$ we obtain
\begin{equation}
\Phi_{||}^{I}=1,\hspace*{1cm}\left(  \frac{\partial\Phi^{I}}{\partial e_{A}%
}\right)  _{||}=\Phi_{||}^{I}\sum_{p}\left(  \frac{1}{\phi_{p}^{I}}%
\frac{\partial\phi_{p}^{I}}{\partial e_{A}}\right)  _{||}=0\label{pa3}%
\end{equation}
and the same result for $\left(  {\partial\Phi}^{I}{/\partial e_{B}}\right)
_{||}$. For the second derivatives we get
\begin{equation}
\left(  \frac{\partial^{2}\Phi^{I}}{\partial e_{A}\partial e_{B}}\right)
_{||}=\left(  \sum_{p}\frac{\partial}{\partial e_{B}}\left(  \frac{1}{\phi
_{p}^{I}}\frac{\partial\phi_{p}^{I}}{\partial e_{A}}\right)  \right)
_{||}=\sum_{p}\frac{\ln^{2}p}{\left(  1-p^{1-ie}\right)  ^{2}}.\label{pa4}%
\end{equation}
In total we get for the contribution to the spectral correlation function
(\ref{paircorrelation})
\begin{equation}
r^{I}(\varepsilon)=-\frac{1}{2\pi^{2}}\mathrm{Re}\left(  \frac{\partial
^{2}Z_{\zeta}^{I}}{\partial e_{A}\partial e_{B}}\right)  _{||}-\frac{1}%
{2\pi^{2}}\mathrm{Re}\left(  \frac{\partial^{2}\Phi^{I}}{\partial
e_{A}\partial e_{B}}\right)  _{||}.%
\end{equation}
Taking into account the relations (\ref{pa1},\ref{pa2},\ref{pa3},\ref{pa4}),
we confirm that this result is identical to the \textquotedblleft
diagonal\textquotedblright\ one obtained by Bogomolny from the Hardy-Littlewood conjecture in \cite{Bog}.

The second contribution to the spectral correlation function, termed
conventionally off-diagonal but which we obtain remaining in the framework of
uncorrelated primes, is provided by the second derivative of $Z^{II}$ with
respect to $e_{A}$ and $e_{B}$ and columnwise identification. This
identification implies that $e_{A}-e_{C}$ and $e_{B}-e_{D}$ go to zero, hence
the following expansion for the inverse zeta functions entering $Z_{\zeta
}^{II}$ can be approximated by
\begin{equation}
\zeta^{-1}(1-i(e_{A}-e_{C}))\zeta^{-1}(1-i(e_{B}-e_{D}))\approx-(e_{A}%
-e_{C})(e_{B}-e_{D})
\end{equation}
such that the only term surviving within this identification results from
$\partial^{2}Z_{\zeta}^{II}/\partial e_{A}\partial e_{B}$ yielding
$-|\zeta(1-ie)|^{2}$. We get in total for the oscillatory (\textquotedblleft
off-diagonal\textquotedblright) contribution to the correlator,
\begin{equation}
r^{II}(e)=\frac{1}{2\pi^{2}}|\zeta(1-ie)|^{2}\mathrm{Re}\left[  \mathrm{e}%
^{2\pi i\bar{\varrho}e}\prod_{p}\left(  1-\frac{(1-p^{ie})^{2}}{(p-1)^{2}%
}\right)  \right]  ,
\end{equation}
where the product in the last equation results from $\left(  \Phi^{II}\right)
_{\parallel}=\Phi^{I}\left(  e/2,e/2,-e/2,-e/2\right)  $ with $\Phi^{I}$ given
by Eq.\ (\ref{PhiI}). Again we reproduced the result of Bogomolny and Keating
obtained in Ref.\ \cite{BogK3} in a heuristic way.

\subsection{$L$-functions}

$L$-functions are generalizations of the Riemann zeta function
\begin{equation}
L(s)=\sum_{n=1}^{\infty}\frac{\chi(n)}{n^{s}}=\prod_{p}\frac{1}{1-\frac
{\chi(p)}{p^{s}}}\label{Ls}%
\end{equation}
including the Dirichlet character $\chi(n)$ defined as a function on integers
periodic with the integer period $k$ such that $\chi(n+k)=\chi(n)$,
multiplicative $\chi(n)\chi(m)=\chi(nm)$, equal to zero iff the greatest
common divisor of $n$ and $k$ is greater than one. We restrict our considerations to real characters
$\chi(n)=\pm1,0$. The generalized Riemann
hypothesis states that all non-trivial zeros of $L(s)$ lie on the critical
line $s=1/2+iE$. The correlation function of these zeros can be determined
using the generating function in a similar way as for the Riemann zeta
function starting from the product representation given in Eq.\ (\ref{Ls}).
The generating function obtained under the assumption ${\rm{Im}}\,%
e_{A,B,C,D}>0$ differs from the Riemann zeta case by the presence of
characters in the factors $z_{p}$; these are now%
\[
z_{p}=\frac{\left(  1-\chi(p)a_{p}\mathrm{e}^{iE\ln p}\right)  \left(
1-\chi(p)b_{p}\mathrm{e}^{-iE\ln p}\right)  }{\left(  1-\chi(p)c_{p}%
\mathrm{e}^{iE\ln p}\right)  \left(  1-\chi(p)d_{p}\mathrm{e}^{-iE\ln
p}\right)  }.
\]
As before, we introduce the diagonal approximation; taking into account that
$\chi^{2}(p)=1,0$, we obtain that $\left\langle z_{p}\right\rangle $ is
given by Eq.\ (\ref{zpav}) if $p$ is not a divisor of $k$, otherwise it drops
out from the product $Z^{I}=\prod_{p}\left\langle z_{p}\right\rangle $.
This implies that the part of the generating function obtained under the
assumption ${\rm{Im}}\,e_{C,D}>0$ can be expressed as
\[
Z^{I}=Z_{\zeta}^{I}(e_{A},e_{B},e_{C},e_{D})\Phi^{I}(e_{A},e_{B},e_{C}%
,e_{D})\Xi^{I}(e_{A},e_{B},e_{C},e_{D})
\]
with $Z_{\zeta}^{I}$ and $\Phi^{I}$ are the same as in Eqs.\ (\ref{zetz}%
,\ref{phip}). The function $\Xi^{I}$ cancels the factors now missing in
$Z^{I}$, it can be expressed as a finite product over the prime divisors of $k
$
\begin{equation}
\Xi^{I}(e_{A},e_{B},e_{C},e_{D})=\prod_{p/k}\frac{1}{\left\langle z_{p}%
\right\rangle }=\prod_{p/k}\left(  1+\frac{(b_{p}-d_{p})(a_{p}-c_{p}%
)}{1-c_{p}d_{p}}\right)  ^{-1}.%
\end{equation}
Differentiation of $Z^{I}$ with respect to $e_{A}$ and $e_{B}$ creates within
the $(\parallel)-$ identification the non-oscillatory contribution to the
correlation function as given in Ref.\ \cite{BogK4}. The oscillatory part is
obtained as $\left(  \partial_{e_{A}e_{B}}^{2}Z^{II}\right)  _{\parallel}$
with $Z^{II}(e_{A},e_{B},e_{C},e_{D})=Z^{I}(e_{A},e_{B},-e_{D},-e_{C});$ it
contains the factor additional to the Riemann zeta case
\begin{equation}
\Psi^{\mathrm{off}}(e)\equiv\left.  \Xi^{I}(e_{A},e_{B},-e_{D},-e_{C})\right\vert
_{\parallel}=\prod_{p/k}\left(  1+\frac{(p^{ie/2}-p^{-ie/2})^{2}}{p-p^{-ie}%
}\right)  ^{-1}%
\end{equation}
which was earlier found in Ref.\ \cite{BogK4} based on the Hardy-Littlewood conjecture. We observe that the correlation
function depends only on the period $k$ but not on the precise form of the
characters $\chi(n)$.

\section{Dynamical systems without time reversal; non-universal corrections}
In this section we return to dynamical systems with classically chaotic counterpart.
We show how the results from Ref.\ \cite{BogK3} can be obtained  by the
method proposed here including the nonuniversal effect of repetitions.

The expression for the spectral determinant $\det(E-H)$ taking into account repetitions of periodic orbits is given by \cite{Haake,Keating}
\begin{equation}
 \det(E-H)=B(E){\rm e}^{-i\bar{N}(E)}\prod_p\prod_{k=0}^\infty\left(1-\frac{{\rm e}^{iS_p(E)/\hbar}}{\Lambda_p^{k+1/2}}\right)
\end{equation}
expressed as product over the primitive periodic orbits $p$ with the actions $S_p(E)$ and the stability coefficients $\Lambda_p={\rm e}^{\lambda_p T_p}$ depending on the Lyapunov exponent $\lambda_p$
and the duration of $T_p$ of
the orbit $p$. The function $\bar{N}(E)$ is given in Eq.\ (\ref{count}) and $B(E)$ again is a real function resulting from the regularization of the infinite product.

Expanding again $S_p(E+e)\approx S_p(E)+e T_p$, $z_p$ is obtained in diagonal approximation as
\begin{equation}\label{zpavn}
 \left\langle z_p\right\rangle=\frac{1}{2\pi}\int_0^{2\pi}d\phi \frac{\left({\rm e}^{ie_C T_p+i\phi}/\sqrt{\Lambda_p};\Lambda_p^{-1}\right)_\infty}{\left({\rm e}^{ie_A T_p+i\phi}/\sqrt{\Lambda_p};\Lambda_p^{-1}\right)_\infty}
 \frac{\left({\rm e}^{ie_D T_p-i\phi}/\sqrt{\Lambda_p};\Lambda_p^{-1}\right)_\infty}{\left({\rm e}^{ie_B T_p-i\phi}/\sqrt{\Lambda_p};\Lambda_p^{-1}\right)_\infty}
\end{equation}
with the $q$-Pochhammer symbols defined as
\begin{equation}
 (x;q)_n=\prod_{k=0}^{n-1}\left(1-xq^k\right).
\end{equation}
The averaging of (\ref{zpavn}) with respect to ${\rm e}^{i\phi}$ is performed by the $q$-binomial formula of Gauss \cite{qbin}
\begin{equation}
 \frac{\left(tx;q\right)_\infty}{\left(x;q\right)_\infty}=\sum_{n=0}^\infty\frac{\left(t;q\right)_n}{\left(q;q\right)_n}x^n
\end{equation}
yielding
\begin{eqnarray}
\fl \left\langle z_p\right\rangle&=&\frac{1}{2\pi}\int_0^{2\pi}\!\!\!d\phi\!\sum_{n,m=0}^\infty\!\!\!\frac{\left({\rm e}^{i(e_C-e_A)T_p};\Lambda_p^{-1}\right)_n}{\left(\Lambda_p^{-1};\Lambda_p^{-1}\right)_n}\!
 \left(\frac{{\rm e}^{i\phi+ie_AT_p}}{\sqrt{\Lambda_p}}\right)^n\!
\frac{\left({\rm e}^{i(e_D-e_B)T_p};\Lambda_p^{-1}\right)_m}{\left(\Lambda_p^{-1};\Lambda_p^{-1}\right)_m}\!\left(\frac{{\rm e}^{-i\phi+ie_BT_p}}{\sqrt{\Lambda_p}}\right)^m\nonumber\\ \fl
&=&\sum_{n=0}^\infty\frac{\left({\rm e}^{i(e_C-e_A)T_p};\Lambda_p^{-1}\right)_n}{\left(\Lambda_p^{-1};\Lambda_p^{-1}\right)_n}
\frac{\left({\rm e}^{i(e_D-e_B)T_p};\Lambda_p^{-1}\right)_n}{\left(\Lambda_p^{-1};\Lambda_p^{-1}\right)_n}\left(\frac{{\rm e}^{i(e_A+e_B)T_p}}{\Lambda_p}\right)^n,
\end{eqnarray}
where the last sum is the definition of the $q$-hypergeometric function of Heine $_2\varphi_1({\rm e}^{i(e_C-e_A)T_p},{\rm e}^{i(e_D-e_B)T_p};1/\Lambda_p;1/\Lambda_p;
{\rm e}^{i(e_A+e_B)T_p}/\Lambda_p)$ \cite{heinehyp,heinehyp1}. In the same way as before in Eq.\ (\ref{zetz}) we can factorize from $Z^I=\prod_p\left\langle z_p\right\rangle$ its poles and zeros
such that the rest is a convergent product. Therefore we consider the inverse classical zeta functions
\begin{equation}
 \frac{1}{Z_{\rm cl}(s)}=\prod_p\frac{1}{Z_{{\rm cl},p}(s)}=\prod_p\prod_{k=0}^\infty\left(1-\frac{{\rm e}^{sT_p}}{\Lambda_p^{k+1}}\right)^{k+1}.
\end{equation}
The component $Z^I$ of the  generating function can then be written as
\begin{eqnarray}
\fl Z^I&=&\frac{Z_{\rm cl}(i(e_A+e_B))Z_{\rm cl}(i(e_C+e_D))}{Z_{\rm cl}(i(e_A+e_D))Z_{\rm cl}(i(e_B+e_C))}
 {\prod_p}\, \left(\frac{Z_{{\rm cl},p}(i(e_A+e_B))Z_{{\rm cl},p}(i(e_C+e_D))}{Z_{{\rm cl},p}(i(e_A+e_D))Z_{{\rm cl},p}
 (i(e_B+e_C))}\right)^{-1}\nonumber\\ \fl &&\times _2\hspace*{-1mm}\varphi_1({\rm e}^{i(e_C-e_A)T_p},{\rm e}^{i(e_D-e_B)T_p};1/\Lambda_p;1/\Lambda_p;
{\rm e}^{i(e_A+e_B)T_p}/\Lambda_p){\rm e}^{-i\pi\overline{\rho}\left(e_C-e_B-e_A+e_D\right)}.
\end{eqnarray}
For $e_{A,B,C,D}\rightarrow 0$
\begin{equation}
\frac{1}{Z_{\rm cl}(i(e_A+e_D))Z_{\rm cl}(i(e_B+e_C))}
\rightarrow-(e_A+e_D)(e_B+e_C)T_0^2
\end{equation}
with a certain reference time $T_0$ \cite{Haake}. The overall generating function is the sum $Z^I+Z^{II}$ where $Z^{II}$
is obtained by the Weyl substitution $e_{C}\rightleftarrows-e_{D}$.
Then the spectral correlation function of \cite{BogK3} is  obtained by Eq.\ (\ref{ZtoR}).

\section{Systems with time reversal}

If time reversal is allowed almost every periodic orbit exists in two versions
with different sense of traversal and exactly the same action. Repeating the
calculations of Section \ref{diagappro} we get the generating function in the diagonal approximation as
$Z=Z_{diag}^{I}+Z_{diag}^{II}$, where%
\begin{eqnarray}
Z_{\mathrm{diag}}^{I}\left(  e_{A,}e_{B},e_{C},e_{D}\right)  & =e^{-i\pi\left(  e_{C}+e_{D}-e_{A}-e_{B}\right)
\bar{\rho}}\frac{\left(  e_{A}+e_{D}\right)  ^{2}\left(  e_{B}+e_{C}\right)
^{2}}{\left(  e_{A}+e_{B}\right)  ^{2}\left(  e_{C}+e_{D}\right)  ^{2}}%
,\nonumber\\
Z_{\mathrm{diag}}^{II}\left(  e_{A,}e_{B},e_{C},e_{D}\right)   &
=Z_{\mathrm{diag}}^{I}\left(  e_{A,}e_{B},-e_{D},-e_{C}\right).\nonumber
\end{eqnarray}
The term $Z_{\mathrm{diag}}^{II}\propto\left(  e_{A}-e_{C}\right)  ^{2}\left(
e_{B}-e_{D}\right)  ^{2}$ does not contribute to the correlation function
since its second mixed derivative by $e_{A,B}$ turns into zero in the
columnwise limit. Therefore in the presence of the time reversal the
oscillatory contribution to the correlation function is a truly off-diagonal
effect different from the Hardy-Littlewood-like mechanism; attempts to obtain
it in the framework of the diagonal approximation were unsuccessful
\cite{BogK3}.

The off-diagonal corrections to the generating function of the orthogonal case
were obtained in \cite{Heusler,Muller}; the symplectic case was considered in
\cite{Braun}. Their existence is due to the so called orbit partnership
existing in the chaotic motion whose significance was realized after the
discovery of the Sieber-Richter pairs \cite{Sieber}. The result has the form
of an asymptotic expansion valid for large scaled energy deviations
$\varepsilon_{X}=e_{X}2\pi\bar{\rho},\quad X=A,B,C,D$. Here the $e_{X}$ are
assumed to be small in classical terms while the dimensionless variables $\varepsilon_{X}$ are considered to be large. In
the orthogonal case the generating function can be written as
$Z_{\mathrm{ortho}}=Z^{I}+Z^{II}$ with%
\begin{eqnarray}
Z^{I}  & =Z_{\mathrm{diag}}^{I}\left(  1+Z_{off}^{I}\right)  ,\nonumber\\
Z_{\mathrm{diag}}^{I}  & =e^{i\left(  \varepsilon_{A}+\varepsilon
_{B}-\varepsilon_{C}-\varepsilon_{D}\right)  /2}\frac{\left(  \varepsilon
_{A}+\varepsilon_{D}\right)  ^{2}\left(  \varepsilon_{B}+\varepsilon
_{C}\right)  ^{2}}{\left(  \varepsilon_{A}+\varepsilon_{B}\right)  ^{2}\left(
\varepsilon_{C}+\varepsilon_{D}\right)  ^{2}},\nonumber\\
Z_{\mathrm{off}}^{I}  & \sim-\frac{\left(  \varepsilon_{A}-\varepsilon
_{C}\right)  \left(  \varepsilon_{B}-\varepsilon_{D}\right)  }{\left(
\varepsilon_{A}+\varepsilon_{D}\right)  \left(  \varepsilon_{B}+\varepsilon
_{C}\right)  }\sum_{n=1}^{\infty}\frac{\left(  -2i\right)  ^{n}\left(
n-1\right)  !}{\left(  \varepsilon_{A}+\varepsilon_{B}\right)  ^{n-1}}\left(
\frac{1}{\varepsilon_{C}+\varepsilon_{D}}+\frac{n}{\varepsilon_{A}%
+\varepsilon_{B}}\right)\nonumber
\end{eqnarray}
and $Z^{II}\left(  \varepsilon_{A},\varepsilon_{B},\varepsilon_{C}%
,\varepsilon_{D}\right)  =Z^{II}\left(  \varepsilon_{A},\varepsilon
_{B},-\varepsilon_{D},-\varepsilon_{C}\right)  .$ The corresponding expansion
in the symplectic case is obtained by the substitution,%
\[
Z_{\mathrm{sympl}}\left(  \varepsilon_{A},\varepsilon_{B},\varepsilon
_{C},\varepsilon_{D}\right)  =Z_{\mathrm{ortho}}\left(  -\varepsilon
_{C},-\varepsilon_{D},-\varepsilon_{A},-\varepsilon_{B}\right)  .
\]
Differentiating the generating functions by $\varepsilon_{A,B}$ and then
setting all arguments equal to $\varepsilon$ we get the correlation
functions as a diverging series in $\varepsilon^{-1}$ which can be brought to
a closed form by the Borel summation. This is essentially a term-by-term Fourier transform to the
time domain followed by summation of the resulting converging series. Finally 
the inverse Fourier transform brings the result back to the
energy domain. An outstanding feature of the symplectic case is that the
asymptotic expansion of the resulting correlation function contains two
oscillatory components proportional respectively to $e^{i2\varepsilon}$ and
$e^{i\varepsilon}$, the latter not present in the input of the Borel
summation; for explanation see \cite{Braun} .

\section{Conclusion}

We present here a new way to compute spectral correlation functions
starting from semiclassical expressions. It is carried out
via the auxiliary generating function which is an averaged ratio of four
spectral determinants. The semiclassical asymptotics of the generating
function consists of two terms responsible for the monotonic and oscillatory
components of the correlator. Our new way is to explain this twofold structure by
complexifying the arguments of the generating functions and bringing them to
the boundaries of convergence of the Gutzwiller periodic orbit expansions of
the spectral determinants. That could be done in two ways differing by the
direction in which we move away from the real axis; in the example of the
Riemann zeta zeros it would mean bringing the arguments of zetas in the
numerator of the generating function either to the left or to the right border
of the critical strip. In our approach after the energy averaging the two semiclassical
approximations of the generating function become well-behaved analytical
functions dominant in the respective complex half-planes; at the real axis the
asymptotics of the generating function is given by their sum.
We avoid the use of the Riemann-Siegel-lookalike formula for the spectral
determinants, so far not proven beyond the Riemann zeta and quantum maps in
finite Hilbert space. We avoid thus the appearance of additional terms caused 
by the sharp cut off at $T_H/2$ and disappearing in the final result. Our approach has
the additional advantage of simplicity, e.g., our derivation of the correlation function
of the Riemann zeta and $L$-functions zeros is probably the simplest existing.

In the absence of the time reversal the oscillatory contribution to the
correlation function can be found via the diagonal approximation to the
generating function totally neglecting the periodic orbit correlations. On the
other hand, if we choose to directly calculate the correlator via the periodic
orbit sum the oscillatory components would exist only if the action
correlation between the periodic orbits is taken into account \cite{Argaman};
in the case of the Riemann zeta zeros this is the correlation of primes
following from the Hardy-Littlewood conjecture. It is still not quite clear
what is the nature of that correlation in physical systems and why it does not
reveal its presence if time reversal is allowed. It is also not clear why the
existence of that correlation automatically follows from the Weyl symmetry of
the generating function.

\section*{Acknowledgements}

We are grateful to F.\ Haake for remarks which stimulated the work on the subject. We also thank T.\ Guhr who read the manuscript and made useful comments
and M.\ Akila, M.\ Bruckhoff for discussions.


\section*{References:}
\begin{thebibliography}{10}
\bibitem{Stockmann} H.-J. St\"{o}ckmann, \textit{Quantum chaos: an introduction}, Cambridge University Press, Cambridge (2008).
\bibitem{Haake} F. Haake, \textit{Quantum Signatures of Chaos}, Springer-Verlag, Berlin, (2010).
\bibitem{Gutzwiller} M.C. Gutzwiller, \textit{Chaos in classical and quantum mechanics}, Springer (1990).
\bibitem{Guhr} T. Guhr, A. M\"{u}ller-Groeling, H. Weidenm\"{u}ller, \textit{Random Matrix Theories in Quantum Physics: Common Concepts}, Phys. Rep. \textbf{299}, 189 (1998).
\bibitem{BGS} O. Bohigas, M.J. Giannoni, C. Schmit, \textit{Characterization of Chaotic Quantum Spectra and Universality of Level Fluctuation Laws}, Phys. Rev. Lett. \textbf{52}, 1 (1984).
\bibitem{Berry} M.V. Berry, \textit{Semiclassical Theory of Spectral Rigidity}, Proc. R. Soc. London, \textbf{400}, 229 (1985).
\bibitem{BogK3} E.B. Bogomolny, J.P. Keating, \textit{Gutzwiller's Trace Formula and Spectral Statistics:
Beyond the Diagonal Approximation}, Phys. Rev. Lett. {\bf77}, 1472 (1996).
\bibitem{Sieber} M. Sieber, K. Richter, \textit{Correlations between periodic orbits and their r\^ole
in spectral statistics}, Physica Scripta, {\bf 2001}, 128 (2001).
\bibitem{Bolte} J. Bolte, S. Keppeler, \textit{Semiclassical Form Factor for Chaotic Systems with Spin 1/2}, J. Phys. A \textbf{32}, 8863 (1999).
\bibitem{Heuslerspin} S. Heusler, \textit{The semiclassical origin of the logarithmic singularity in the symplectic form factor}, J. Phys. A \textbf{34}, L483 (2001).
\bibitem{BolteHar} J. Bolte, J. Harrison, \textit{The spin contribution to form factor of quantum graphs}, J. Phys. A \textbf{36}, 2747 (2003).
\bibitem{MullerPhysRev} S. M\"{u}ller, S. Heusler, P. Braun, F. Haake, A. Altland, \textit{Periodic-orbit theory of universality in quantum chaos}, Phys. Rev. E \textbf{72}, 046207 (2005).
\bibitem{Heusler} S. Heusler, S. M\"{u}ller, A. Altland, P. Braun, F. Haake, \textit{Periodic-Orbit Theory of Level Correlations}, Phys. Rev. Lett. \textbf{98}, 044103 (2007).
\bibitem{Muller} S. M\"{u}ller, S. Heusler, A. Altland, P. Braun, F. Haake, \textit{Periodic-orbit theory of universal level correlations in quantum chaos}, New J. Phys. \textbf{11}, 103025 (2009).
\bibitem{Braun} P. Braun, \textit{Beyond the Heisenberg time: semiclassical treatment of spectral correlations in chaotic systems with spin 1/2}, J.\ Phys.\ A {\bf45}, 045102 (2012).
\bibitem{NagaoMuller} T. Nagao, S. M\"{u}ller, \textit{
The n-level spectral correlations for chaotic systems}, J. Phys. A \textbf{42}, 375102, (2009).
\bibitem{Keatin} J.P.\ Keating, \textit{The semiclassical functional equation}, Chaos {\bf2}, 15 (1992).
\bibitem{KeaMul} J.P.\ Keating, S.\ M\"uller, \textit{Resummation and the semiclassical theory of spectral statistics}, Proc.\ R.\ Soc.\ A {\bf463}, 3241 (2007).
\bibitem{Wal} D.\ Waltner, S.\ Gnutzmann, G.\ Tanner, K.\ Richter, 
\textit{Subdeterminant approach for pseudo-orbit expansions of spectral determinants in quantum maps and quantum graphs}, Phys.\ Rev.\ E {\bf87}, 052919 (2013).
\bibitem{Har} R.\ Band, J.M.\ Harrison, C.H.\ Joyner, \textit{Finite pseudo orbit expansions for spectral quantities of quantum graphs}, J.\ Phys.\ A {\bf45}, 325204 (2012).
\bibitem{Tit} E.C.\ Titchmarsh, \textit{The Theory of the Riemann Zeta-function}, Oxford University Press, Oxford (1988).
\bibitem{Edw} H.M.\ Edwards, \textit{Riemann's Zeta Function}, Academic, New York, 1974.
\bibitem{Ber500} M.V.\ Berry, \textit{The Riemann-Siegel expansion for the zeta function: high orders and remainders}, Proc.\ R.\ Soc.\ Lond. A {\bf{450}}, 439462 (1995).
\bibitem{Bog} E.B. Bogomolny, \textit{Riemann Zeta Function and Quantum Chaos}, Progress of Theo. Phys.
Supp., {\bf166} 19 (2007).
\bibitem{Keating} J.P. Keating, \textit{The Riemann Zeta-Function and Quantum Chaology.} in Quantum Chaos, G. Casati,
        I. Guarneri, U. Smilansky, eds., North-Holland, Amsterdam, pp. 145-185, (1993).
\bibitem{Argaman} N. Argaman, F.-M. Dittes, E. Doron, J.P. Keating, A.Yu. Kitaev, M. Sieber, U. Smilansky, \textit{Correlations in the Actions of Periodic Orbits Derived from Quantum Chaos},
Phys. Rev. Lett. \textbf{71}, 4326 (1993).
\bibitem{Montgomery} H.L. Montgomery, \textit{The pair correlation of zeros of the zeta function}, Analytic Number Theory (St. Louis, 1972), Proc. Sympos. Pure Math. \textbf{24}, Amer. Math. Soc. (Providence), pp. 181-193, (1973).
\bibitem{Odlyzko} A.M. Odlyzko, \textit{The $10^{20}$-th zero of the Riemann zeta function an 70 million of its neighbors.}
        Preprint, AT\& T Bell Laboratories, (1989).
\bibitem{BogK1} E.B. Bogomolny, J.P. Keating, \textit{Random matrix theory and the Riemann zeros
I: three and four-point correlations}, Nonlinearity \textbf{8}, 1115 (1995).
\bibitem{BogK2} E.B. Bogomolny, J.P. Keating, \textit{Random matrix theory and the Riemann zeros
II: n-point correlations}, Nonlinearity \textbf{9}, 911 (1996).
\bibitem{BogK4} E.B. Bogomolny, J.P. Keating, \textit{Two-point correlation function for Dirichlet
$L$-functions}, J.\ Phys.\ A {\bf46}, 095202 (2013).
\bibitem{BogK5} E.B. Bogomolny, J.P. Keating, \textit{A method for calculating spectral statistics based on
random-matrix universality with an application to the three-point correlations of the Riemann zeros},
J.\ Phys. A, {\bf46}, 305203 (2013).
\bibitem{BerK} M.V. Berry, J.P. Keating, \textit{The Riemann zeros and eigenvalue asymptotics},
SIAM Review \textbf{41}, 236 (1999).
\bibitem{Conrey} J. B. Conrey, N. C. Snaith, \textit{Applications of the $L$-Functions Ratios Conjectures}, Proc.\ London Math. Soc. {\bf 94}, 594 (2007).
\bibitem{qbin} Wolfram Math World: q-Binomial Theorem, \textit{http://mathworld.wolfram.com/q-BinomialTheorem.html}.
\bibitem{heinehyp} Wolfram Math World: q-Hypergeometric Function, \textit{http://mathworld.wolfram.com/q-HypergeometricFunction.html}.
\bibitem{heinehyp1} E.\ Heine, Untersuchungen \"uber die Reihe \textit{$1+\frac{\left(1-q^\alpha\right)\left(1-q^\beta\right)}{\left(1-q\right)\left(1-q^\gamma\right)}x
    +\frac{\left(1-q^\alpha\right)\left(1-q^{\alpha+1}\right)\left(1-q^\beta\right)\left(1-q^{\beta+1}\right)}{\left(1-q\right)
    \left(1-q^2\right)\left(1-q^\gamma\right)\left(1-q^{\gamma+1}\right)}x^2+\ldots$}, J.\ reine angew.\ Math.\ {\bf34}, 285 (1847).
    \end{thebibliography}
\end{document}